\documentclass[namedreferences]{solarphysics}
\usepackage[hyperref,optionalrh,solaromanenum]{spr-sola-addons} 
\usepackage{graphicx}                    
\usepackage{color}                       
\usepackage{breakurl}                         

\begin{document}
\sloppy
\begin{article}

\begin{opening}

\title{Positive and Negative Photospheric Fields in Solar Cycles 21 - 24}

%
\author[addressref={aff1},corref,email={helena@ev13934.spb.edu}]{\inits{E.S.}\fnm{E.S.}\lnm{Vernova}\orcid{0000-0001-8075-1522}}
\author[addressref=aff1]{\inits{M.I.}\fnm{M.I.}~\lnm{Tyasto}}\sep
\author[addressref=aff2]{\inits{D.G.}\fnm{D.G.}~\lnm{Baranov}\orcid{0000-0003-2838-8513}}\sep
\author[addressref=aff1]{\inits{O.A.}\fnm{O.A.}~\lnm{Danilova}}\sep

%
\runningauthor{E.S.~Vernova \textit{et al.}}
\runningtitle{Positive and Negative Photospheric Fields}

\address[id={aff1}]{IZMIRAN, SPb. Filial, Laboratory of Magnetospheric
Disturbances, St. Petersburg, Russia}
\address[id={aff2}]{Ioffe Institute, St. Petersburg, Russia}
\begin{abstract}
Distribution of the positive and negative photospheric  fields is
studied on the base of synoptic maps of the photospheric magnetic
field produced by the National Solar Observatory Kitt Peak (NSO
Kitt Peak) for $1976-2016$. In the analysis only the sign of the
field irrespective of its strength is taken into account,
emphasizing the role of the fields with average and weak
strengths. Time changes in the positive and negative magnetic
fields for two hemispheres, for the high latitudes and for the
sunspot zone  as well as their imbalances are considered.
Distributions of fields of opposite polarities, which are defined
mostly by the fields of the high latitudes, change with a 22-year
cycle. The polarity imbalance in  each hemisphere is closely
connected with the dipole moment $g_{10}$. The imbalance of high
latitudes for two hemispheres changes with the 22-year period and
coincides with the sign of both the polar field in the southern
hemisphere, and the quadrupole moment $g_{20}$. For the magnetic
fields of the sunspot zone during $\sim75\%$ of time a connection
of the magnetic field imbalance with the quadrupole moment taken
with the reversed sign $- g_{20}$ and with the sign of the polar
field in the northern hemisphere is observed. The received results
testify to cyclic changes in the polarity imbalance.
\end{abstract}

%

\end{opening}


\section{Introduction}
All variety of manifestations of the solar activity (SA) is
related to magnetic fields of various strength and their
localization on the surface of the Sun. The distribution of
magnetic fields over the surface of the Sun and its change in the
course of the solar cycle is one of the key points in creating of
the solar dynamo models (see, for example, \opencite{char10}).
Evolution of zonal distribution of the Sun's magnetic field was
considered by \inlinecite{hoek91} on the basis of magnetograms of
the Wilcox Solar Observatory (WSO). It was shown that the magnetic
flux is closely connected to the level of activity and is similar
to the Maunder butterfly diagram which reflects the distribution
of sunspots. Variations of magnetic fields in time and latitude
are considered by \inlinecite{akht15} where the strong difference
of variations of the magnetic flux at low and high latitudes is
revealed.

In contrast to the 11-year solar cycle which shows itself in the
increase and following decrease of the number and area of
sunspots, the 22-year recurrence of  the magnetic field is
observed as changes in polarity of  the solar magnetic fields:
sign change of the leading and following sunspots in the solar
cycle minimum (Hale's law), sign change of the polar field near SA
maximum. Thus, the polarity of magnetic fields of the Sun is
important characteristic of many processes, which are connected
with the recurrence of SA.

Field lines of the magnetic field are closed, therefore the
magnetic fluxes of the Sun of the positive and negative polarity
have to be equal. However by consideration of separate
manifestations of SA the balance of polarity is often broken. For
example, the imbalance of magnetic fields of the leading and
following sunspots is a well-known fact. In Cycle 21  at the
northern/southern active regions the positive/negative fields were
on average stronger, than negative/positive (\opencite{petr12}).
During Cycle 22 the situation was opposite, and the picture
returned again to initial when Cycle 23  began. These relations
reflect widely known fact that magnetic fields of bipolar groups
of sunspots display the asymmetry with  the leading polarity being
as a rule stronger.

The imbalance is observed also in the polar fields which are
antisymmetric except for some abnormal periods. During the
reversal of the polar field the two hemispheres develop rather
independently, therefore polar fields do not complete the change
of sign reversals synchronously (\opencite{sval13}). In some
periods the global field of the Sun lost the dipole character and
looked as a monopole (\opencite{wilc72, koto09}).

The dipole character of the global field is manifested in the
defining role of the dipole moment among coefficients of the
magnetic field multipole expansion according to the PFSS model
(the potential-field source-surface model (\opencite{hoek86})).
The dominating terms of the expansion are the dipole, hexapole,
and  quadrupole, the main field components during minima of SA
being the axial ones (\opencite{brav00}). While the dipole and
hexapole have opposite  polarities at the Sun's poles, the
quadrupole has the same sign at both poles. In each minimum of SA
the polarity of the quadrupole is opposite to the dipole and the
hexapole polarity in the north and coincides with the polarity of
the dipole and the hexapole in the south. Because of  this
property the quadrupole moment increases the polar magnetic field
of the southern hemisphere, at the same time weakening the polar
field of the northern hemisphere which results in asymmetry of the
magnetic field of the Sun. Weak magnetic fields occupy most of the
solar surface. \inlinecite{iosh09} showed that small-scale
background magnetic fields  form a population which has a tendency
to specific cyclic variations.

Exploring the imbalance of the positive and negative magnetic
fields we take into consideration only the sign of the field,
irrespective of the field strength. Due to this approach the
fields with the weak and average strength are observed more
distinctly, because only the surface area occupied with fields of
certain polarity was taken into account. The purpose of this
research was studying of the distribution of the photospheric
fields with opposite polarities over the solar surface and the
change of this distribution during a 22-year magnetic cycle.

\begin{figure}[h]
\begin{center}
\includegraphics[width=0.9\textwidth]{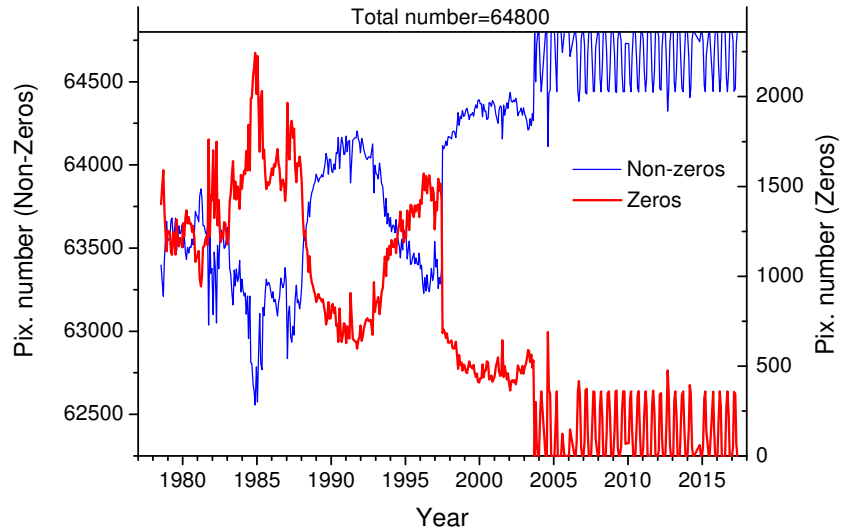}
\caption{The number of zero (red line) and non-zero (blue line)
values of the magnetic field strength in synoptic maps (number of
pixels). Each map consists of 64800 pixels. Data of KPVT:
$1978-2003$; data of SOLIS: $2003-2016$. } \label{zeros}
\end{center}
\end{figure}

\section{Data and method}

We used as basic data synoptic maps of the photospheric magnetic
field of National Solar Observatory Kitt Peak (NSO Kitt Peak).
Combining the data obtained by two instruments of the Kitt Peak
Observatory - KPVT (Kitt Peak Vacuum Telescope) for $1976-2003$
(\url{ftp://nispdata.nso.edu/kpvt/synoptic/mag/}) and SOLIS/VSM
(Synoptic Optical Long-term Investigations of the Sun, Vector
Solar Magnetometer) for $2003-2016$
(\url{https://magmap.nso.edu/solis/archive.html}) we studied the
change of positive and negative magnetic fields during four solar
cycles. Synoptic maps of the photospheric magnetic field are
produced with the resolution of $1^\circ$ of the longitude (360
steps) and 180 equal steps in the sine of the latitude. Thus, each
map contained 360x180 pixels of magnetic field in gauss. Because
of large gaps in data during the initial stage of the KPVT
operation interval $1976-1977$ was not considered.

A color scale displaying the strength of the magnetic field is
often used while creating diagrams latitude vs. time on the basis
of synoptic maps. Such a scale usually represents the strength of
the magnetic field from the weakest to strongest fields of the
sunspot zone. As a result strong fields are most distinctly shown
in diagram, reproducing only the  distribution of sunspots
(Maunder butterfly diagram) and suppressing fields with lower
strength. We set another task: to consider mainly distribution of
fields with average and low strength. Replacing value of the field
strength B in each of pixels with +1 for $B>0$, and with -1 for
$B<0$, we consider only the sign of the field, disregarding
strength value. Thus, we estimate the area occupied with fields of
a certain polarity while sunspots occupying a small part of the
solar surface give only insignificant contribution. The fields
with $B>100$\,G make about $1\%$ of pixels while the fields with
$100> B> 5$\,G occupy $34\%$ of the solar surface, and the fields
with $B <5$\,G - $65\%$. We considered distribution of fields
$B>0$ and $B<0$, having excluded pixels with $B=0$.

The number of pixels with the non-zero field strength $|B|> 0$ and
the number of pixels with $B=0$ presented in Figure~\ref{zeros}
show that zero values of the strength appear not so much as a
result of measurements but rather as a result of gaps in data. An
unexpected jump of zeros number (Figure~\ref{zeros}) occurred in
1997 which is perhaps connected with change of the instrument
sensitivity. In the dataset of SOLIS, since the second half of
2003, not all lines of pixels near to the poles are filled by
extrapolation. This explains the emergence of a peculiar "comb" -
fluctuations of zeros number. In general, pixels with $B=0$ make a
small percent (no more $\sim 3.5\%$) from the total number of
pixels in the synoptic map, and they can be neglected. Those
rotations in which there were considerable gaps in data (number of
zero values more than $3.5\%$) were not  used in the analysis. 4
rotations were excluded from the dataset for $1978-2003$, 15
rotations were excluded from the dataset of SOLIS for $2003-2016$.

\section{Results and discussion}
\begin{figure}[h]
\begin{center}
\includegraphics[height=5.5 cm]{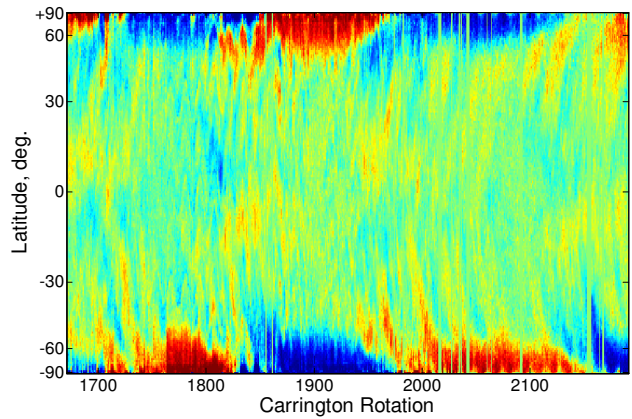}
\caption{The diagram latitude-time for Carrington rotations
$1670-2190$ ($1978-2016$). The color scale represents the relative
abundance of the positive and negative pixels at each latitude
(red color corresponds to the presence only of positive pixels $B>
0$\,G, blue color corresponds to the presence only of negative
pixels $B <0$\,G).} \label{map}
\end{center}
\end{figure}

We replace the values of the magnetic field strength with $+1$ or
$-1$ according to the sign of the field in a pixel and average
each synoptic map over longitude. In this way the single
latitude-time diagram for $1978-2016$ (Carrington rotations from
1670 to 2190) is obtained where the color scale from blue $-1$ to
red $+1$ represents the relative abundance of positive and
negative pixels as a function of latitude and rotation number
(Figure~\ref{map}).

The 22-year periodicity in the magnetic field polarity is shown as
alternation of blue and red regions near the poles. The middle
points of the corresponding time intervals coincide with  the   SA
minima. However the 11-year cycle of SA is hardly distinguishable
in the sunspot zones because the contribution of the strong fields
is suppressed. As a result the alternating blue-green and
red-yellow inclined strips going from the equator to poles became
clearly seen. Apparently, these strips correspond to streams of
the magnetic fields of different polarities which are transferred
to the poles due to the meridional circulation. The inclination of
these strips (number of degrees/rotation) allows to evaluate the
speed of the field transfer. The estimates give the following
speed values: for the northern hemisphere $13.1\pm 0.8$\,m/s and
for southern (where separate streams are more pronounced) $12.9\pm
0.6$\,m/s. These estimates are slightly lower than the value of
speed 20\,m/s obtained by measurement of Doppler shift (see
\opencite{petr12} and references in it), however they lie within
the limits of $10 - 20$\,m/s observed by \inlinecite{roud18}.

\begin{figure}[t]
\begin{center}
\includegraphics[width=0.5\textwidth]{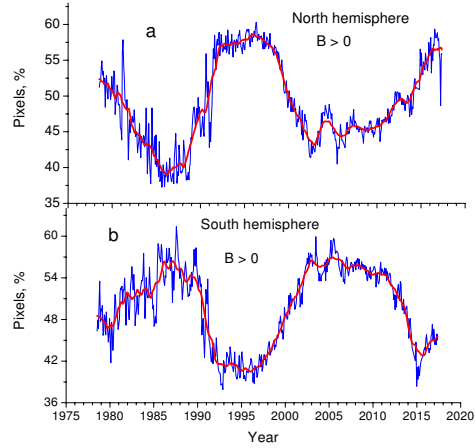}
\caption{The area of the Sun's surface (as a percentage) occupied
with magnetic fields (a) of positive polarity in the whole
northern hemisphere $N_{plus}$ and (b) of positive polarity in the
southern hemisphere $S_{plus}$ for each Carrington rotation. Red
lines show the sliding smoothing on 20 rotations.} \label{hemall}
\end{center}
\end{figure}

Most clearly appear in Figure~\ref{map} the near-polar regions
where one of the two polarities alternately dominates defining the
general dominating polarity of this hemisphere. In
Figure~\ref{hemall}a time change of relative number of pixels with
the positive polarity $N_{plus}$ is presented as a percentage for
the northern hemisphere. It can be seen that the area occupied by
the positive fields in the northern hemisphere is closely
connected with the sign of the polar field in this hemisphere and
changes with the 22-year period. During those periods of time when
the sign of the northern polar field was positive, the area
occupied by the positive polarity in SA minimum is up to $60\%$ of
the hemisphere surface (Figure~\ref{hemall}a). Similar conclusions
apply to the time changes of the negative field percentage, but
the negative fields in the northern hemisphere develop in an
antiphase with the fields of the positive polarity, reaching the
maximum area when the sign of the northern polar field is
negative. In Figure~\ref{hemall}b relative change in number of
pixels with positive polarity $S_{plus}$ is presented as a
percentage for the southern hemisphere. The same conclusions, as
for the northern hemisphere are true: the positive polarity of the
southern hemisphere is closely connected with the sign of the
polar field in the southern hemisphere and reaches the maximum
values $\sim 60\%$ in years close to the SA minimum. The
dominating fields in each hemisphere during 11 years are those
which sign coincides with the sign of the polar field in this
hemisphere. Thus, the imbalance of positive and negative fields in
each hemisphere changes with the 22-year period.

\begin{figure}[h]
\begin{center}
\includegraphics[width=0.5\textwidth]{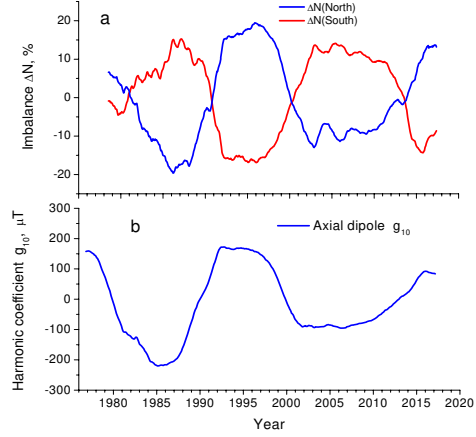}
\caption{(a) The imbalance of the number of positive and negative
pixels as a percentage for the northern and southern hemispheres:
$\Delta N(North)=N_{plus}-N_{minus}$ (blue line) and $\Delta
N(South)=S_{plus}-S_{minus}$ (red line), respectively. (b) axial
dipole moment $g_{10}$ of the magnetic field (WSO).} \label{g10}
\end{center}
\end{figure}
In Figure~\ref{g10}a imbalances of magnetic fields for the
northern hemisphere $N_{plus} - N_{minus}$ and the southern
hemisphere $S_{plus} - S_{minus}$ are presented. Though in this
case we do not take into account the magnetic field strength B and
consider the whole range of latitudes, we receive results close to
those obtained in (\opencite{vern18}) for high latitudes (from
$40^\circ$ up to $90^\circ$) and the fields with $B<50$\,G.
Consequently the dipole magnetic field is the determining factor
for the polarity imbalance in a separate hemisphere. In
Figure~\ref{g10}b the axial dipole moment $g_{10}$ is shown which
was calculated at the Wilcox Solar Observatory (WSO,
\url{http://wso.stanford.edu/}) using PFSS model. The polarity
imbalance in the northern hemisphere changes in phase with the
dipole moment $g_{10}$ (Figure~\ref{g10}a,b).

\begin{figure}[t]
\begin{center}
\includegraphics[width=0.5\textwidth]{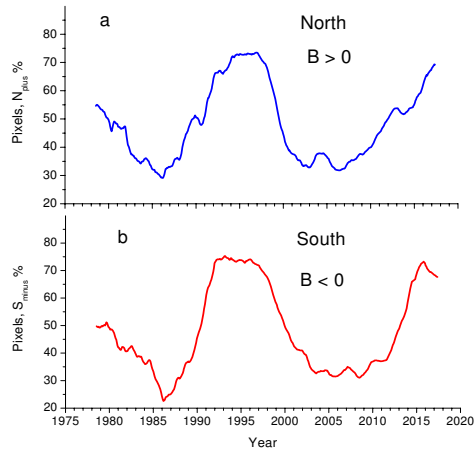}
\caption{High heliolatitudes. The area of the Sun's surface (as a
percentage) occupied with the magnetic fields (a) of positive
polarity in the northern hemisphere  $N_{plus}$ (heliolatitudes
from $+40^\circ$  to $+90^\circ$) and (b) of negative polarity in
the southern hemisphere $S_{minus}$ (from $-40^\circ$  to
$-90^\circ$). Data are averaged over  20 solar rotations. }
\label{high}
\end{center}
\end{figure}

Connection of the surface area occupied by the fields of a certain
polarity with the sign of the polar field is most pronounced for
the high-latitude fields (from $40^\circ$ to $90^\circ$). We
compare positive fields of the northern hemisphere
(Figure~\ref{high}a) and negative fields of the southern
hemisphere (Figure~\ref{high}b). These fields develop in phase
showing the correlation coefficient $R=0.94$. At high latitudes
the fields of the positive polarity occupy about $80\%$ of the
surface when the sign of the northern polar field is positive
during SA minimum (Figure~\ref{high}a). If we consider the whole
hemisphere, the percentage is about $60\%$ (see
Figure~\ref{hemall}a). Two characteristic groups of the
large-scale magnetic fields were described by \inlinecite{andr08}:
the first group represented weak background fields $3 - 10$\,G,
the second group represented stronger fields $75 - 100$\, G. It
was shown that weak fields with the N-polarity in the northern
hemisphere correlate with the S-polarity fields in the southern
hemisphere. This result is in good agreement with
Figure~\ref{high}.
\begin{figure}[h]
\begin{center}
\includegraphics[width=0.6\textwidth]{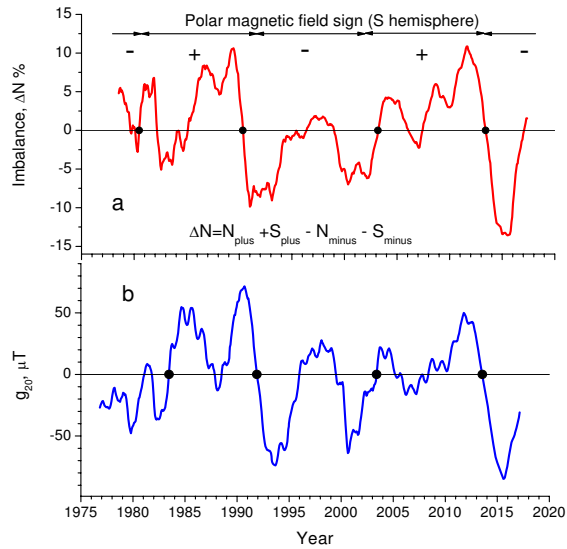}
\caption{The polarity imbalance at the high latitudes. (a)
Heliolatitudes from $+40^\circ$ to $+90^\circ$ and from
$-40^\circ$ to $-90^\circ$: the difference of number of positive
and negative pixels in polar regions of two hemispheres ($\Delta N
= N_{plus} + S_{plus} - N_{minus} - S_{minus}$). In the top part
of the figure the sign of the polar field in the southern
hemisphere is shown; (b) quadrupole moment $g_{20}$ of magnetic
field (WSO). Black circles mark the sign changes of the imbalance
near periods of the polar field reversal.} \label{imbhigh}
\end{center}
\end{figure}

To estimate the total polarity imbalance  for high latitudes
($40^\circ - 90^\circ$) we considered the difference of number of
positive and negative pixels $\Delta N$ in polar regions of both
hemispheres (Figure~\ref{imbhigh}a):
\begin{equation}  \label{dpix}
\Delta N = N_{plus} + S_{plus} - N_{minus} - S_{minus}
\end{equation}

During four solar cycles the strict regularity of the imbalance
change is observed. For 11 years from one solar maximum to another
the imbalance keeps the sign, changing it during the period close
to the reversal of the polar field. Thus the  periodicity of the
imbalance-sign change is 22 years. The sign of the imbalance
coincides with the sign of the polar field in the southern
hemisphere (the sign is shown at the top of
Figure~\ref{imbhigh}a). In Figure~\ref{imbhigh}b the quadrupole
moment $g_{20}$ of the photospheric magnetic field (WSO) is shown.
The polarity imbalance and the quadrupole moment have nearly the
same time course, displaying good coincidence of their signs.
These results agree with our conclusions in (\opencite{vern18})
where the magnetic field strength was taken into account.

\begin{figure}[h]
\begin{center}
\includegraphics[width=0.7\textwidth]{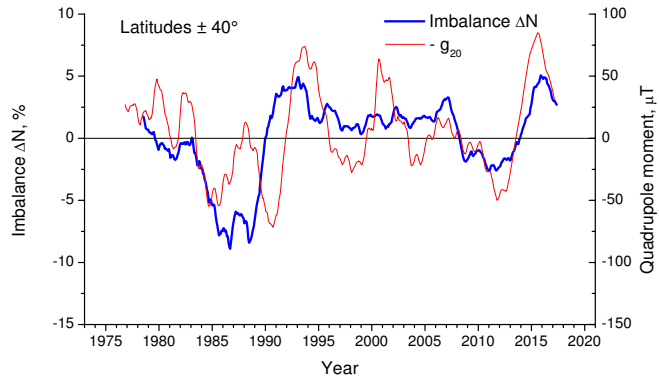}
\caption{The imbalance of the positive and negative pixel number
at the equatorial latitudes $\pm40^\circ$ - the blue line and the
quadrupole moment with the reversed sign ($-g_{20}$)}- the red
line.
 \label{imblow}
\end{center}
\end{figure}

In the sunspot zone the positive and negative fields occupy
approximately equal areas. Nevertheless in the latitude range from
$-40^\circ$ to $+40^\circ$ some difference between the number of
the positive and negative pixels is observed (Figure~\ref{imblow},
the blue line), the maximum of the imbalance being $5-8 \%$ of the
total number of pixels. The quadrupole moment taken with the
reversed sign $-g_{20}$ is shown in Figure~\ref{imblow} (the red
line). For several time periods both parameters change the sign
almost simultaneously. The similarity of changes in the imbalance
and the quadrupole moment is observed from 1978 to 2003 though the
imbalance  is a little shifted relative to the quadrupole moment.
Discrepancy of these curves is seen at the decrease phase of Cycle
23  when the imbalance continues to remain positive (2003--2008)
while the quadrupole moment $-g_{20}$ becomes negative. From 2009
to 2016 the imbalance and the quadrupole moment change almost
synchronously.

The recurrence of the imbalance is similar to the features of the
imbalance of strong magnetic fields (\opencite{vern18}). Change of
the sign of the imbalance in the years close to reversals of the
polar field was the main feature of the imbalance of the strong
fields of the sunspot zone. The sign of the imbalance of the
strong fields changes with the 22-year period just as the sign of
the quadrupole moment $-g_{20}$ and coincides with the sign of the
polar field in the northern hemisphere. The same features can be
seen in Figure~\ref{imblow} in the imbalance of the positive and
negative fields considered irrespective of the magnetic field
strength. Thus, though not for the whole interval of time, there
is a connection of the sign of the magnetic field imbalance of the
sunspot zone with the quadrupole moment $-g_{20}$ and with the
sign of the polar field in the northern hemisphere. The time
interval  when the sign of the imbalance and the sign of the
quadrupole moment coincide  is about $75\%$ of the considered
period of time.

\section{Conclusions}
1. We consider distribution of the positive and negative magnetic
fields over the surface of the Sun for four solar cycles, taking
into account only the field sign irrespective of its strength.
Such approach allows us to emphasize the peculiarities of the
distribution of the magnetic fields with average and small
strengths due to the suppression of the strong fields connected
with  active regions of the Sun. Thus, we study the area occupied
with fields of the positive and negative polarities.

2. We show that the features of the distribution for the  fields
of different polarity are defined generally by the  high-latitude
fields. The polarity imbalance  in each hemisphere is closely
connected with the changes in the dipole moment $g_{10}$. The
effect of the dipole component of the field  on the imbalance  of
the positive and negative fields is observed for the whole
hemisphere, and not just for high latitudes, as it was in the case
when the magnetic field strength was taken into account.

3. For high latitudes the imbalance of positive and negative
fields in the latitude ranges from $+40^\circ$ to $+90^\circ$ and
from $-40^\circ$ to $-90^\circ$ is considered. The sign of the
imbalance changes with the 22-year period and coincides both with
the sign of the polar field in the southern hemisphere, and with
the sign of the quadrupole moment $g_{20}$. These important
results are obtained by taking into account only the area occupied
by the fields of positive or negative polarities.

4. For the most part of time ($\sim 75\%$) for the fields in the
latitude range   $\pm40^\circ$ we observe coincidence of the
imbalance sign of magnetic fields with the reversed sign of
quadrupole moment  $-g_{20}$ and with the sign of the polar field
in the northern hemisphere.

5. The diagram latitude-time shows structures in the form of
inclined strips  which became discernible due to the domination of
one of the polarities. The inclination of these strips,
apparently, is connected with transfer of magnetic fields in the
direction from the equator to the poles. In that case the speed of
transfer will be equal to about 13 m/s which is close to estimates
of speed of meridional circulation made by the other authors.

\section{Acknowledgements}

NSO/Kitt Peak data used here are produced cooperatively by
NSF/NOAO, NASA/GSFC, and NOAA/SEL. This work utilizes SOLIS data
obtained by the NSO Integrated Synoptic Program (NISP), managed by
the National SolarObservatory. Wilcox Solar Observatory data used
in this study was obtained via the web site
http://wso.stanford.edu  courtesy of J.T. Hoeksema.

\end{article}
\end{document}